\documentclass{vgtc}                          

\ifpdf
  \pdfoutput=1\relax                   
  \usepackage{graphicx}                
  \DeclareGraphicsExtensions{.pdf,.png,.jpg,.jpeg} 
\else
  \usepackage{graphicx}                
  \DeclareGraphicsExtensions{.eps}     
\fi%

\graphicspath{{figures/}} 

\usepackage{microtype}                 
\PassOptionsToPackage{warn}{textcomp}  
\usepackage{textcomp}                  
\usepackage{mathptmx}                  
\usepackage{times}                     
\usepackage{cite}                      
\usepackage{tabu}                      
\usepackage{booktabs}                  

\onlineid{1067}

\vgtccategory{Research}

\vgtcinsertpkg

\usepackage{subcaption}
\usepackage{amsmath}

\newcommand{\ac}[1]{{\color{black}#1}}

\title{Simulating the Geometric Growth of the Marine Sponge Crella Incrustans}

\author{Joshua O'Hagan\thanks{e-mail: joshohagan17@outlook.co.nz}\\ %
        \scriptsize Computational Media Innovation Centre,\\ \scriptsize Victoria University of Wellington %
\and Andrew Chalmers\thanks{e-mail: andrew.chalmers@vuw.ac.nz}\\ %
     \scriptsize Computational Media Innovation Centre,\\ \scriptsize Victoria University of Wellington %
\and Taehyun Rhee\thanks{e-mail: taehyun.rhee@vuw.ac.nz}\\ %
     \scriptsize Computational Media Innovation Centre,\\ \scriptsize Victoria University of Wellington}

\teaser{
  \centering
  \includegraphics[width=0.95\linewidth]{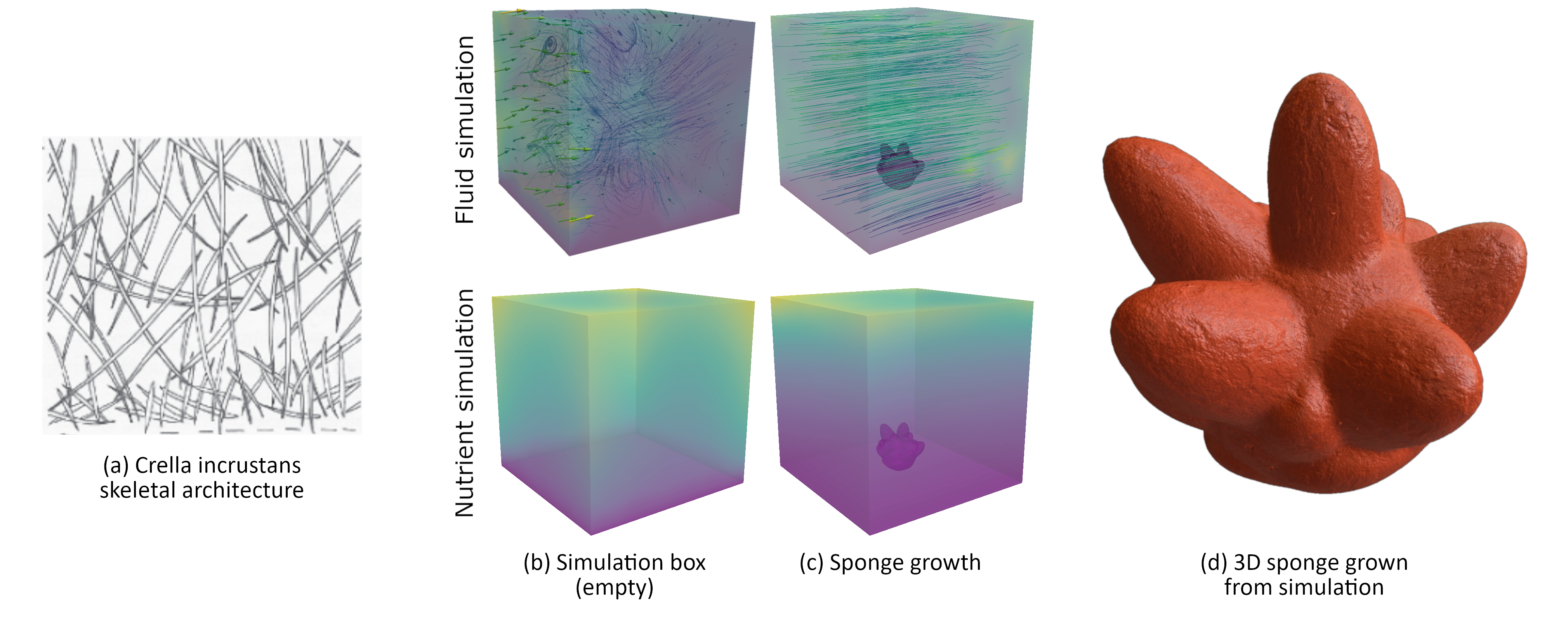}
  \caption{Overview of simulating the marine sponge Crella \ac{incrustans}. Given the (a) skeletal architecture that resembles Crella incrustans and the (b) simulation box for fluid and nutrients, we are able to (c) simulate sponge growth using the skeletal architecture to guide the growth pattern. This results in a (d) 3D \ac{mesh} of Crella incrustans.}
  \label{fig:teaser}
}

\abstract{Simulating marine sponge growth helps marine biologists analyze, measure, and predict the effects that the marine environment has on marine sponges, and vice versa. This paper describes a way to simulate and grow geometric models of the marine sponge Crella incrustans while considering environmental factors including fluid flow and nutrients. The simulation improves upon prior work by changing the skeletal architecture of the sponge in the growth model to better suit the structure of Crella \ac{incrustans}. The change in skeletal architecture and other simulation parameters are then evaluated qualitatively against photos of a real-life Crella incrustans sponge. The results support the hypothesis that changing the skeletal architecture from radiate accretive to Halichondrid produces a sponge model which is closer in resemblance to Crella incrustans than the prior work.}

\CCScatlist{
    \CCScatTwelve{Computing methodologies}{Computer graphics}{Shape modeling}{};
    \CCScatTwelve{Computing methodologies}{Modeling and simulation}{}{};
    \CCScatTwelve{Human-centered computing}{Visu\-al\-iza\-tion}{Scientific visualization}{}
}

\begin{document}


\firstsection{Introduction} \label{C:intro}

\maketitle


Crella incrustans is a marine sponge that is threatened by the recent marine heat waves in New Zealand~\cite{Strano2022}. Such marine sponges are an important part of the marine ecosystem, filtering bacteria and providing habitats to marine organisms~\cite{Strano2021}. Understanding the growth process of marine sponges can help in the protection of marine sponges against climate change. 3D geometric sponge simulation alongside marine environment simulation can be used to efficiently evaluate and forecast the effects that climate change has on marine sponges. 

The problem of geometric growth simulation can be split into two main parts: simulation of the environment, and simulation of the sponge growth. For the simulation of sponge growth, we focus only on the overall shape, not on ostia or small bumps on the surface.

The prior work~\cite{Kaandorp2013IB2013p114} simulates branching sponges (e.g., ``Haliclona oculata'') by simulating the environment through a fluid and nutrient simulation, and simulates the sponge growth through the ``Accretive Growth Model''~\cite{Kaandorp2001p}. The main limitation of the Accretive Growth Model is that the simulation is not able to produce geometric models similar to the sponge Crella incrustans, which can be attributed to the difference in skeletal architecture used in the simulation. 

We hypothesize that changing the skeletal architecture, as well as fine-tuning the simulation parameters, will produce geometric models that will more closely resemble the sponge Crella incrustans. We change the skeletal architecture from Radiate Accretive~\cite{Kaandorp2013IB2013p114} to Halichondrid in the simulation and evaluate the result. We follow the Accretive Growth Model to simulate a sponge depositing silicon layers, represented as triangles, during the growth process. We simulate the ocean fluid flow using the finite element method~\cite{TURNER1956} with the Navier-Stokes equations, and the nutrient distribution simulation with the \ac{finite element method} using the advection-diffusion equations \ac{(\textbf{}Figure~\ref{fig:teaser})}. 
\section{Related Work}\label{C:rel-work}

Early work on marine sponge simulation~\cite{Kaandorp1991p7188} uses fractals to model the growth of the skeleton of the sponge Haliclona oculata in 2D. Transplantation experiments~\cite{Kaandorp1992} were then used to evaluate real-life sponge growth with the 2D fractal model. These models were able to predict real-life phenomena, such as the sponge growing thin-branching plate-like ends when transplanted from a sheltered site to an exposed site. The radiate accretive growth model was later introduced for sponge growth, again focusing on the case study of Haliclona oculata~\cite{Kaandorp1993p4161}. This model adds a nutrient simulation to the fractal techniques~\cite{Kaandorp1991p7188}, where the sponge's growth rate is based on the number of nutrient particles absorbed. The nutrient is simulated by the Diffusion Limited Aggregation model. This model is also extended to 3D in the work, where the lattice Boltzmann method is used to simulate the hydrodynamics of the ocean current~\cite{KAANDORP2000p2441}. The accretive growth model is summarized in by Kaandorp et al.~\cite{Kaandorp2001p}.

Improvements to the accretive growth model were introduced, including a more accurate growth model, fluid velocities of up to 0.05m/s are simulated using the COMSOL Finite Element Method advection-diffusion simulation, and a model of gene regulation~\cite{Kaandorp2013IB2013p114,Chindapol2013}. Speed improvements have also been proposed~\cite{Abela2015} at the cost of fluid and nutrient simulation accuracy.



\section{Crella Incrustans Skeletal Architecture}\label{C:improvements-sim-model}

Sponges can take on different skeletal architectures based on environmental conditions, such as the velocity of the water flow, or the exposure to the water flow~\cite{Uriz2003MRaT62p279299}. The way the spicules are laid out during the sponge growth determines the skeletal architecture. The three main skeletal architectures~\cite{Kaandorp2009} are Radiate accretive, Axinellid, and Halichondrid, all following an accretive growth process where new layers of spicules are deposited onto previous layers, and other spicules connect vertically between the layers. The direction in which the layers of spicules are deposited depends on the type of skeletal architecture. The Halichondrid skeletal architecture has layers of spicule rings deposited in a more randomized set of directions to the surface, which tends to cause encrusting shapes. Whereas the radiate accretive skeletal architectures have spicule rings deposited on the surface of the sponge at an angle normal to the surface, and the Axinellid skeletal architecture has spicule rings deposited in the direction of the central surface normal in an area in the middle of the branch, and in the direction of the surface normal elsewhere. 

\ac{From photos} of Crella incrustans (Figure~\ref{fig:changeArc}a), we obverse they either have a radiate accretive architecture or a Halichondrid architecture. The radiate accretive architecture is already present in the~\cite{Kaandorp2013IB2013p114} model, so we experiment with the Halichondrid architecture in an attempt to reproduce the more encrusting forms of Crella incrustans. \ac{Our simulation model's default parameters are based on the prior work~\cite{Kaandorp2013IB2013p114, Chindapol2013} unless otherwise specified.}

The Halichondrid architecture has spicules placed vertically with randomness. We simulated this randomness by changing the vertex normal of the sponge vertices, $\overrightarrow N_i$, to $\overrightarrow N_i'$, before growing the vertices for each growth iteration of the simulation. $\overrightarrow N_i'$ is calculated as:

\begin{align}
\label{eq:norm-offset}
\overrightarrow N_i' = normalize(\overrightarrow N_i + (x_{rand}, y_{rand}, z_{rand}))
\end{align}

where

\begin{align}
x_{rand}, y_{rand}, z_{rand} \in [-radius, radius]
\end{align}

and $radius$ is a chosen parameter (we use 6cm) determining how much the normals will be skewed. \ac{Each axis ($x_{rand}$, $y_{rand}$, $z_{rand}$) are randomly generated independent from one another.}

\section{Geometric Simulation of Crella incrustans}\label{C:geo-sim-crella-incrustans}

\begin{figure}[th]
    \centering
    \includegraphics[width=\columnwidth]{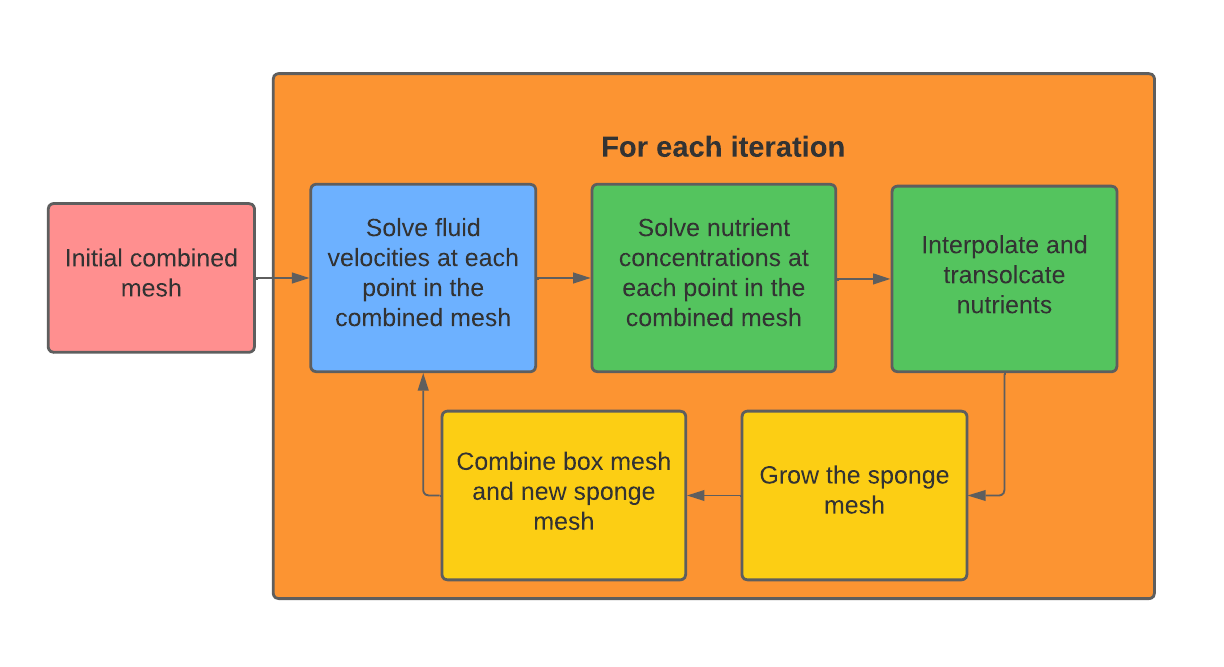}
    \caption{An overview of the sponge simulation.}
    \label{sponge-sim-overview}
\end{figure}

 The simulation consists of an ocean fluid simulation, nutrient distribution simulation, and then a sponge growth step is performed for each iteration (see Figure \ref{sponge-sim-overview}). The simulation repeats the loop until a maximum iterations is reached. The ocean fluid flow is simulated using the finite element method (FEM)~\cite{TURNER1956} with a non-linear equation solver for the Navier-Stokes equations.

\subsection{Mesh Setup}
\begin{figure*}[t!]
    \centering
    \begin{subfigure}[t]{0.3\textwidth}
        \centering
        \includegraphics[width=\textwidth]{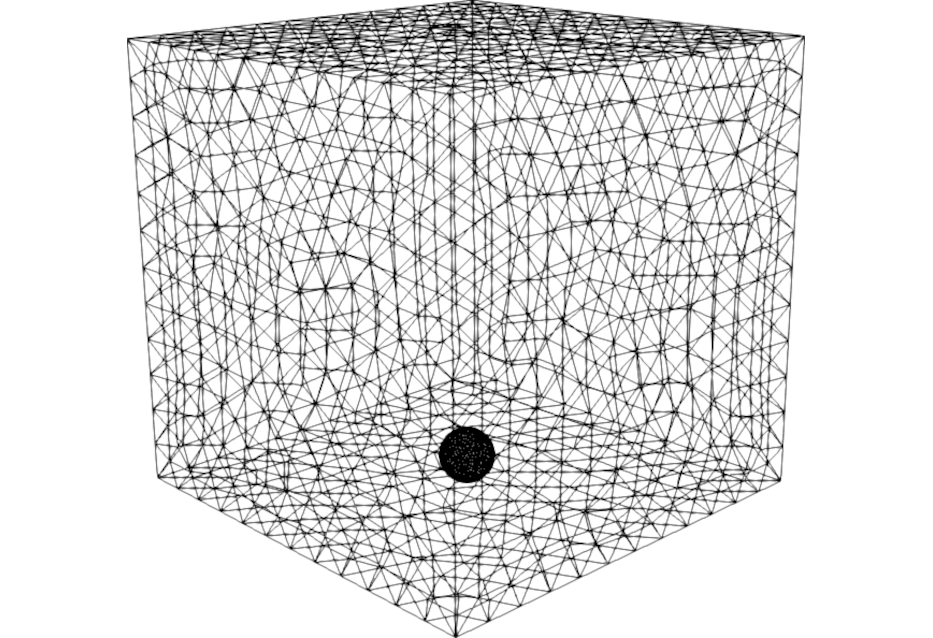}
        \caption{Initialization of the sponge and box mesh combined.}
        \label{fig:mesh-sim-box}
    \end{subfigure}%
    ~ 
    \begin{subfigure}[t]{0.3\textwidth}
        \centering
        \includegraphics[width=\textwidth]{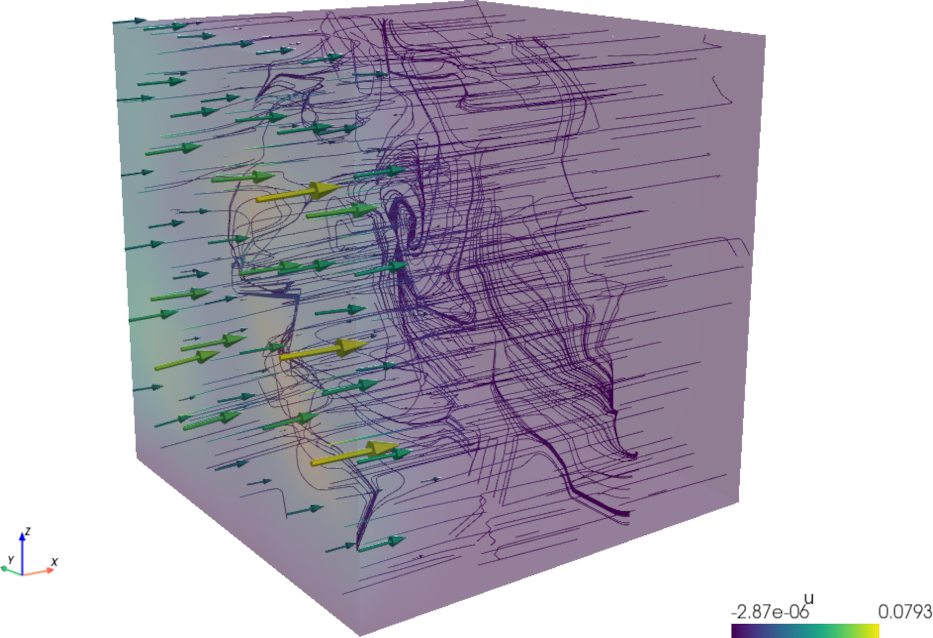}
        \caption{Fluid flow simulation.}
        \label{fig:fluid-sim-box}
    \end{subfigure}%
    ~ 
    \begin{subfigure}[t]{0.3\textwidth}
        \centering
        \includegraphics[width=\textwidth]{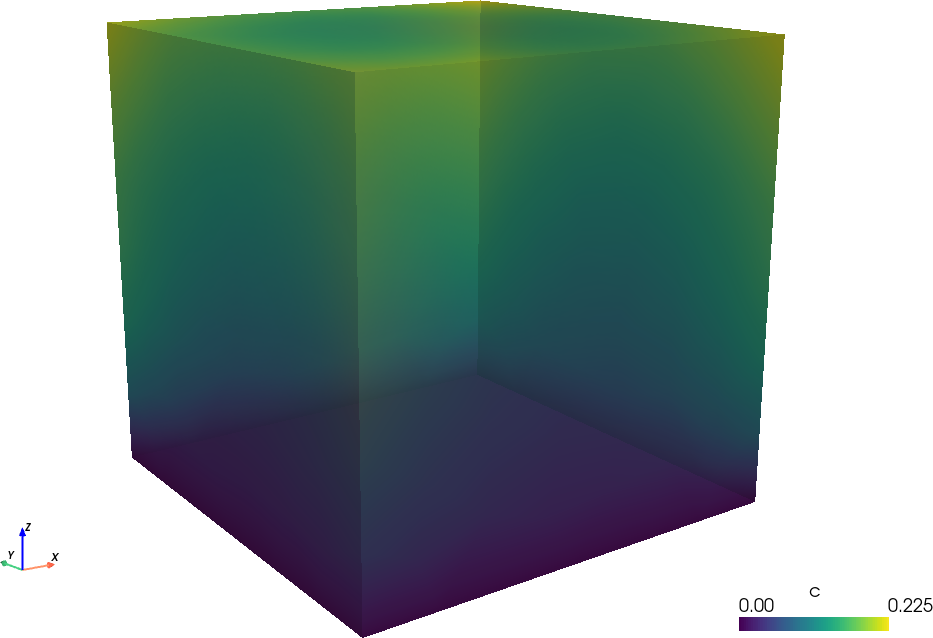}
        \caption{Nutrient simulation.}
        \label{fig:nutrient-sim-box}
    \end{subfigure}
    \caption{Simulation visualization for (a) shape, (b) fluid, and (c) nutrients. The (b) arrows convey the strength of the fluid flow at a given point (through size and color), and particle streamlines are shown in dark blue. The (c) yellow to blue visualizes high to low nutrient concentration respectively.}
\end{figure*}

The first step is to create the box and sponge mesh. The box mesh is an FEM cube tetrahedral mesh, with each cell split in half each time the resolution is increased (starting from a resolution of 1). A higher-resolution box mesh is good for having more points where the nutrient is simulated, which can help create more branches in the final sponge model. A tetrahedral mesh is commonly used for FEM problems when there is complex fluid to solid boundaries. The sponge mesh is initialized as an icosphere of radius 6cm \ac{(based on prior work for comparison~\cite{Kaandorp2013IB2013p114}, which represents a small sponge before it begins growth)}, with a given subdivision value \ac{determining the resolution (Figure~\ref{fig:mesh-sim-box}).}

\subsection{FEM Fluid Equation Solver}

After the sponge and box mesh are initialized, the fluid flow field $u$, in the incompressible Navier-Stokes equations (NVS) and the continuity equation, is solved for on the mesh using the FEM (see Figure \ref{fig:fluid-sim-box}).

The Navier-stokes equations are defined as:

\begin{align}
\label{eq:nvs}
\rho \frac{\partial \overrightarrow u}{\partial t} 
- \nabla \cdot[\eta (\nabla \overrightarrow u + \nabla \overrightarrow u ^{T})] 
+ \rho (\overrightarrow u \cdot \nabla) \overrightarrow u
+ \nabla p
= F, 
\\
\label{eq:continuity}
\nabla \cdot \overrightarrow u
= 0,
\end{align}

where $F$ is the external volume force (set to zero), $u$ is the fluid velocity, $p$ is the pressure, $\eta$ is the dynamic viscosity, and $\eta (\nabla \overrightarrow u + \nabla \overrightarrow u ^{T})$ is the stress tensor resulting from the fluid viscosity, and $t$ is the time step.

The initial condition of the fluid velocity is set to $0.05m^{-s}$ along the inlet (left-hand side of the box along the positive x-axis). The substratum and surface of the sponge have boundary conditions set to the no-slip condition, where $u = \overrightarrow 0$. The pressure, $p$, and viscous stress, $\eta (\nabla \overrightarrow u + \nabla \overrightarrow u ^{T}) n$, is set to zero along the outlet, which specifies vanishing viscous stress, and a Dirichlet condition on the pressure. The rest of the boundaries are open with no viscous stress - $\eta (\nabla \overrightarrow u + \nabla \overrightarrow u ^{T}) n = 0$, which means that fluid is free to enter or leave the simulation domain along these boundaries.~\cite{Chindapol2013} suppresses turbulence by increasing the viscous forces over inertial forces due to instabilities in the simulation. This means that the fluid viscosity is increased to $5.0e^{-2} Pa s$, which is about $50$ times more than water. We apply the same concept in our simulation.

The stationary, time-independent solutions are then solved for (meaning that $\frac{\partial \overrightarrow u}{\partial t} = \overrightarrow 0$)~\cite{Cimrman2019AiCM45p18971921}), which uses the non-linear Newton solver. The relative tolerance of the Newton solver is set to $1e-3$, and the max iterations is set to $100$. For the linear solver used by the Newton solver, we set the relative tolerance is set to $1e-6$, and the max iterations is set to $100$. 

\subsection{Nutrient Distribution Simulation}
For the nutrient distribution simulation, we use the FEM again (Figure~\ref{fig:nutrient-sim-box}). Having the nutrient simulation use the same technique as the ocean fluid flow simulation - the FEM - makes the overall simulation much simpler and more consistent. A non-linear equation solver with the Advection-Diffusion equations is used on an FEM mesh, where the fluid velocity is taken as input to the Advection-Diffusion equations. The non-linear equation solver solves the advection-diffusion equation

\begin{equation}
\label{eq:advec-diff}
\frac{\partial c}{\partial t}
+ \overrightarrow u \cdot \nabla c
= D \nabla^{2} c
\end{equation}

where $c$ is the concentration gradient, $t$ is the time step, $u$ is the fluid velocity vector solved in the NVS and continuity equations, and $D$ is the diffusion coefficient.

In order to pass the fluid velocity to the advection-diffusion solver, the fluid velocities must be provided for each point in the FEM mesh. Therefore, the fluid velocities are ``probed'' (i.e. interpolated).

An initial condition of the advection-diffusion solver is that all wall boundaries (except the substratum) are initialized to the idealized value of $1.0 mol m^{-3}$. The boundary condition of $0$ nutrient concentration is set on the sponge and substratum boundary, as nutrient is constantly absorbed by these boundaries in real life.

The relative tolerance of the Newton solver is set to $1e-6$, and the max iterations is set to $25$. The linear solver's relative tolerance is set to $1e-6$, and the max iterations is set to $100$.

\subsection{Sponge Growth}
\begin{figure*}[th]
    \centering
    \includegraphics[width=0.2\textwidth]{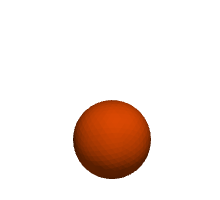}
    \label{sponge_growth_1}
    \includegraphics[width=0.2\textwidth]{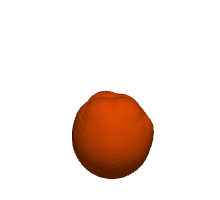}
    \label{sponge_growth_2}
    \includegraphics[width=0.2\textwidth]{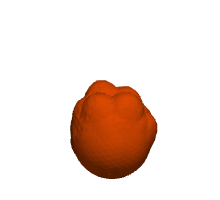}
    \label{sponge_growth_3}
    \includegraphics[width=0.2\textwidth]{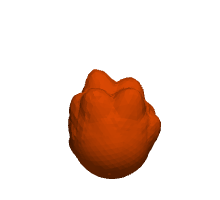}
    \label{sponge_growth_4}
    \caption{A simulation of sponge growth at iterations 1, 15, 30, and 45 respectively.}
    \label{sponge-growth}
\end{figure*}

The thickness of the new skeleton layer during sponge growth is linearly related to the amount of nutrient absorbed~\cite{Kaandorp2001p}. In order to determine the distance a sponge vertex will grow when forming the next skeleton layer, the nutrient absorbed at each vertex on the surface of the sponge is calculated. Rather than adding a new layer of vertices for the formation of the next skeleton layer, we found it more efficient and straightforward to displace the current vertices based on their growth distance. Additionally, vertex insertion and fusion rules are performed during the formation of the next skeleton layer to ensure the edges of triangles in the mesh continue to approximate the distance between two spicule rings~\cite{Kaandorp2013IB2013p114}. See Figure~\ref{sponge-growth} for an example of simulating sponge growth.

\subsection{Sponge Surface Nutrient Concentrations}
Because the nutrient will always be zero at the sponge boundary (the sponge vertices), the nutrient concentration for a given sponge vertex must be calculated slightly offset to the surface. Therefore, the concentration is calculated by probing nutrient concentrations, which are specified at the points in the box mesh, along the normal of the vertex (scaled by a given offset distance). An attenuation function is then applied to the concentrations, where d represents the concentration value's distance along the normal (Equation \ref{eq:atten}). This is done to ensure that closer nutrient values have more influence on sponge growth, as sponges in the ocean are more likely to absorb closer nutrients than nutrients that are further away. After attenuation, the nutrient values along the normal are averaged to give the concentration value absorbed by the given sponge vertex:

\begin{equation}
\label{eq:atten}
    c(d) = \frac{1}{20 d + 1}, d \in [0, 1],
\end{equation}

\ac{where the $20$ was experimentally chosen to scale up the attenuation of concentration values.} After the nutrient concentration is calculated at each sponge vertex, the nutrient is translocated across the surface of the sponge in our simulation through means of surface diffusion. Nutrient translocation is a biological process that occurs in corals, not sponges, but can be used in our sponge simulation to simulate a more spread out branching pattern. The translocation for each vertex $x$, and each neighbouring vertex $x'$, is approximated using the following equation:

\begin{align}
    c_{tr}(x') = clamp(c(x') + T_{give} c(x)) \\
    c_{tr}(x) = clamp(c(x) - N_{neigh} T_{give} c(x))
\end{align}

where $c_{tr}(x')$ represents the updated nutrient concentration at a neighbouring vertex $x'$, $c_{tr}(x)$ represents the updated nutrient concentration at the current vertex, $T_{give} \in [0, 1]$ is the coefficient determining how much nutrients to give to the neighbours, $N_{neigh}$ is the number of neighbouring vertices, and $clamp$ is a function which clamps the value to the range $[0, 1]$. 

For each vertex in the sponge mesh, translocation is performed on the given vertex, and then on each neighbour of the given vertex, so that the nutrient spreads over a large range of the sponge surface, giving more spread out and realistic branches.

If the amount of nutrient on a portion of a sponge's surface is below a certain threshold, the gene Iroquois will be produced~\cite{Kaandorp2013IB2013p114}. Iroquois causes the portion of the sponge's surface to stop growing and begin to form the aquiferous system. We simulate this process by taking the nutrient value of a given vertex after translocation and only allowing the vertex to grow if its nutrient value is above a chosen threshold.

The growth function~\cite{Chindapol2013} is then used to determine the growth length:

\begin{equation}
    \label{eq:chin}
    l_i = g(C_{i}) \overrightarrow N_i
\end{equation}

\begin{equation}
    g(C_i) = clampMG(Gr * L_{max} \frac{C_i^n}{K + C_i^n})
\end{equation}

where $l_i$ is the growth length of a given vertex $i$, $C_i$ is the translocated nutrient concentration at vertex $i$, $n$ is the kinetic order of the growth rate with respect to $C_i$ and $\overrightarrow N_i$ is the vertex normal at vertex $i$. $Gr$ is the growth rate, chosen so that we can better control the sponge growth rate, and $clampMG$ is a function that clamps the value to the range $[0, max growth]$, where $max growth$ is a chosen value which allows us to avoid too much sponge growth in one iteration. $L_{max}$ is an asymptotic maximum growth rate, and $K$ is used with $L_{max}$ to denote the characteristic growth curve of the sponge.

The final growth vector, $l_i$, is then added to the vertex position, and the result is clamped to be inside the simulation box. This results in the new, displaced, vertex position representing the sponge growth.

\subsection{Vertex Insertion and Fusion}

We can model the distance between two sponge spicule rings as the edges between triangle vertices in our initial spherical mesh~\cite{Kaandorp2013IB2013p114}. As the sponge mesh grows, vertices will naturally get further and further away from each other due to the spherical mesh normals pointing away from each other. This means that the edges will become larger, no longer representing a realistic distance between spicule rings in real-life sponge biology. To overcome this problem, we insert a vertex when the edge is longer than a given threshold, and new edges and triangles are formed based on a set of rules~\cite{Merks2003}. \ac{The new vertex is inserted along} the two-dimensional polynomial formed by the two vertices. After vertex insertion, vertices can also move too close together, which now means that the distances between spicule rings are too small. To overcome this problem, vertices are also fused based on a distance threshold, and again a set of rules is used to define how the new edges and triangles are formed~\cite{Merks2003}. The rules used cover all possible insertion and fusion scenarios in the triangle mesh.

\subsection{Crella Incrustans Sponge Growth Randomization}
The FEM simulation always returns the exact same nutrient values, given the same set of parameters, so the exact same sponge shape is produced at each growth step. This is not ideal for the purposes of simulating the biological randomness in real life that causes every sponge of a given species to grow into its own unique shape. To better replicate Crella incrustans, We randomize the position of the sponge on the x and y axes within the simulation. This allows the sponge to be close to different points providing different nutrient concentrations to be absorbed during the sponge simulation.

\section{Evaluation and Results}\label{C:eval-and-results}

\begin{figure}[!t]
    \centering
    \includegraphics[width=\columnwidth]{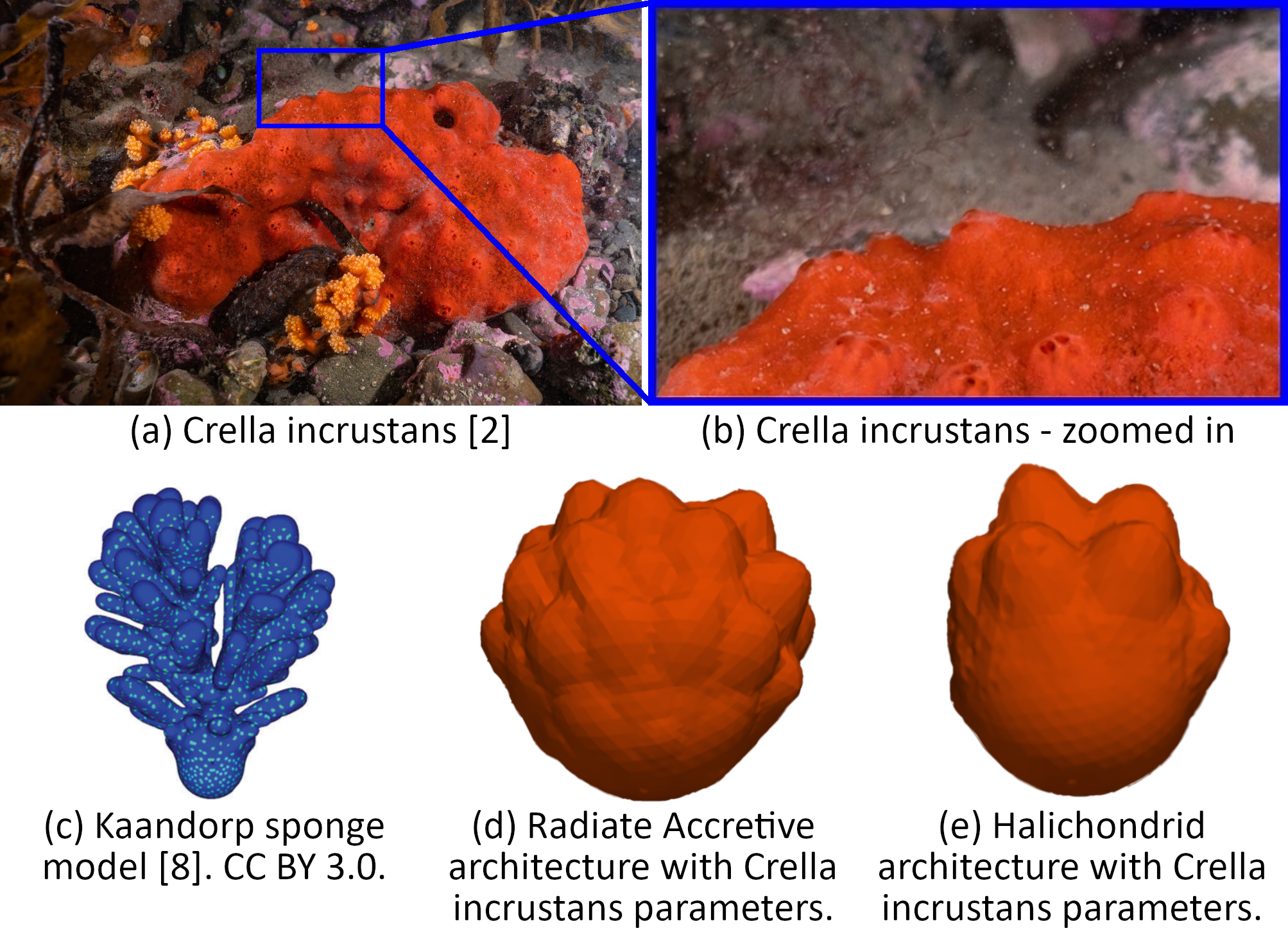}
    \caption{\ac{Comparing (a, b) a real-life photo~\cite{Bell2022} with (e) the simulated Crella incrustans sponge. We also compare with prior work (c) as well as demonstrate changing the skeletal architecture from (d) Radiate Accretive to (e) Halichondrid to better simulate Crella incrustans.}}
    \label{fig:changeArc}
\end{figure}

The performance of our sponge simulation is evaluated by qualitatively measuring the time taken for 50 growth iterations under different sets of simulation parameters. The sponge growth result is shown in Figure~\ref{sponge-growth}. We show a comparison with a photo \ac{and prior work} in Figure~\ref{fig:changeArc}. \ac{The prior skeletal architecture used in Figure~\ref{fig:changeArc}d is similar in shape to the top of the cropped photo of Crella incrustans, but the extrusions are too dense, too close together, and not tall enough. Our proposed architecture (Figure~\ref{fig:changeArc}e produces less dense, more spread out, and taller extrusions, better resembling the shapes produced by Crella incrustans. Additional results for varying the parameters and photo comparisons are in the supplementary materials. This also includes a table of the simulation parameters and their descriptions.}

\section{Conclusion}\label{C:con}

This paper describes a geometric sponge simulator that can produce a variety of different sponges by changing the simulation parameters. Changes in the skeletal architecture were made in order to better simulate the sponge Crella incrustans. A qualitative evaluation was performed to test the effectiveness of changing the skeletal architecture. The results show that changing the skeletal architecture produces a sponge that closely resembles Crella incrustans.

A limitation of this work is the time and effort in tuning the paramters, where each simulated sponge shown in the results took about 2 hours. Future work could explore automating this process. Future work could also explore integrating additional parameters to closely match the effects of climate change.

\acknowledgments{The authors thank Prof. James Bell, Dr.
Valerio Micaroni, and Francesca Strano for their insights on sponge biology. This work was supported by the Entrepreneurial University Programme from the TEC and the Smart Ideas Endeavour Fund from MBIE, New Zealand.}

\bibliographystyle{abbrv-doi}

\bibliography{ms}
\end{document}



\firstsection{Supplementary} 

\maketitle

\begin{figure*}[t!]
    \centering
    \begin{subfigure}[t]{0.2\textwidth}
        \centering
        \includegraphics[width=\textwidth]{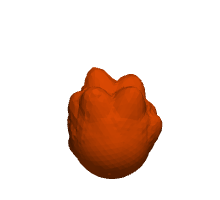}
        \caption{0.1 (default)}
        \label{fig:c-offset-0-1}
    \end{subfigure}%
    ~ 
    \begin{subfigure}[t]{0.2\textwidth}
        \centering
        \includegraphics[width=\textwidth]{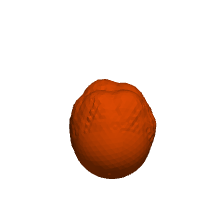}
        \caption{0.2.}
        \label{fig:c-offset-0-2}
    \end{subfigure}%
    ~ 
    \begin{subfigure}[t]{0.2\textwidth}
        \centering
        \includegraphics[width=\textwidth]{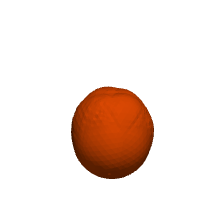}
        \caption{0.3.}
        \label{fig:c-offset-0-3}
    \end{subfigure}
    ~ 
    \begin{subfigure}[t]{0.2\textwidth}
        \centering
        \includegraphics[width=\textwidth]{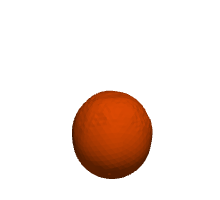}
        \caption{0.4.}
        \label{fig:c-offset-0-4}
    \end{subfigure}
    \caption{Changing the \textit{concentration offset distance} simulation parameter. This parameter determines how far along the normal concentration values are interpolated.}\label{fig:c-offset}
\end{figure*}

\begin{figure*}[t!]
    \centering
    \begin{subfigure}[t]{0.2\textwidth}
        \centering
        \includegraphics[width=\textwidth]{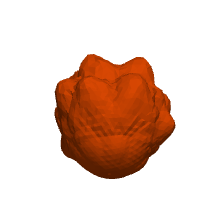}
        \caption{1.0.}
        \label{fig:chindapol-n-1-0}
    \end{subfigure}%
    ~ 
    \begin{subfigure}[t]{0.2\textwidth}
        \centering
        \includegraphics[width=\textwidth]{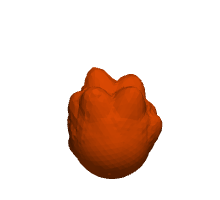}
        \caption{1.4 (default).}
        \label{fig:chindapol-n-1-4}
    \end{subfigure}%
    ~ 
    \begin{subfigure}[t]{0.2\textwidth}
        \centering
        \includegraphics[width=\textwidth]{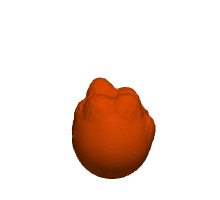}
        \caption{1.8.}
        \label{fig:chindapol-n-1-8}
    \end{subfigure}
    ~ 
    \begin{subfigure}[t]{0.2\textwidth}
        \centering
        \includegraphics[width=\textwidth]{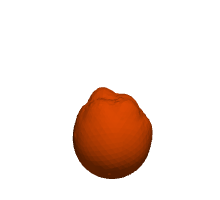}
        \caption{2.2.}
        \label{fig:chindapol-n-2-2}
    \end{subfigure}
    \caption{Changing the \textit{Chindapol kinetic order} simulation parameter. This parameter determines how much the sponge spreads out.}
    \label{fig:chin-n}
\end{figure*}

\begin{figure*}[t!]
    \centering
    \begin{subfigure}[t]{0.2\textwidth}
        \centering
        \includegraphics[width=\textwidth]{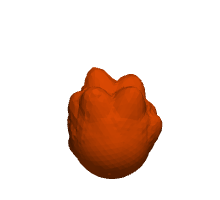}
        \caption{1.0 (default).}
        \label{fig:chindapol-n-2-2-2}
    \end{subfigure}%
    ~ 
    \begin{subfigure}[t]{0.2\textwidth}
        \centering
        \includegraphics[width=\textwidth]{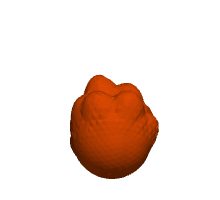}
        \caption{1.4.}
        \label{fig:chindapol-n-1-0-2}
    \end{subfigure}%
    ~ 
    \begin{subfigure}[t]{0.2\textwidth}
        \centering
        \includegraphics[width=\textwidth]{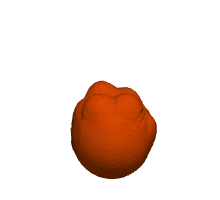}
        \caption{1.8.}
        \label{fig:chindapol-n-1-4-2}
    \end{subfigure}
    ~ 
    \begin{subfigure}[t]{0.2\textwidth}
        \centering
        \includegraphics[width=\textwidth]{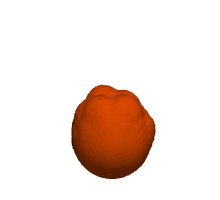}
        \caption{2.2.}
        \label{fig:chindapol-n-1-8-2}
    \end{subfigure}
    \caption{Changing the \textit{Chindapol characteristic growth constant} simulation parameter. This parameter controls the asymptotic growth curve.}
    \label{fig:chin-K}
\end{figure*}

\begin{figure*}[t!]
    \centering
    \begin{subfigure}[t]{0.2\textwidth}
        \centering
        \includegraphics[width=\textwidth]{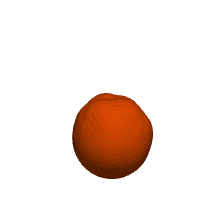}
        \caption{1.25.}
        \label{fig:chindapol-n-2-2-2}
    \end{subfigure}%
    ~ 
    \begin{subfigure}[t]{0.2\textwidth}
        \centering
        \includegraphics[width=\textwidth]{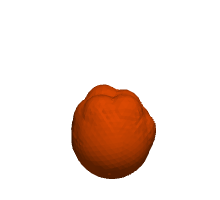}
        \caption{2.5.}
        \label{fig:chindapol-n-1-0-2}
    \end{subfigure}%
    ~ 
    \begin{subfigure}[t]{0.2\textwidth}
        \centering
        \includegraphics[width=\textwidth]{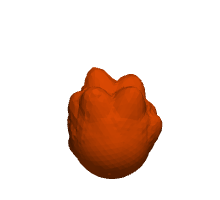}
        \caption{5 (default).}
        \label{fig:chindapol-n-1-4-2}
    \end{subfigure}
    ~ 
    \begin{subfigure}[t]{0.2\textwidth}
        \centering
        \includegraphics[width=\textwidth]{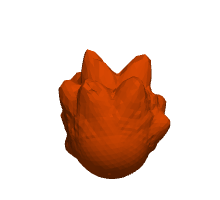}
        \caption{10.}
        \label{fig:chindapol-n-1-8-2}
    \end{subfigure}
    \caption{Changing the \textit{growth rate} simulation parameter. The growth rate determines how much the sponge grows for each iteration.}
    \label{fig:growth-rate}
\end{figure*}

\begin{figure*}[t!]
    \centering
    \begin{subfigure}[t]{0.2\textwidth}
        \centering
        \includegraphics[width=\textwidth]{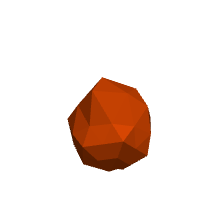}
        \caption{$1$}
        \label{fig:chindapol-n-1-0-2}
    \end{subfigure}
    \begin{subfigure}[t]{0.2\textwidth}
        \centering
        \includegraphics[width=\textwidth]{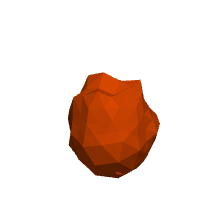}
        \caption{$2$}
        \label{fig:chindapol-n-1-4-2}
    \end{subfigure}
    \begin{subfigure}[t]{0.2\textwidth}
        \centering
        \includegraphics[width=\textwidth]{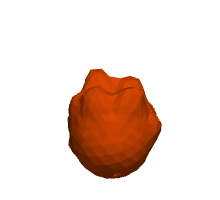}
        \caption{$3$}
        \label{fig:chindapol-n-1-8-2}
    \end{subfigure}
    \begin{subfigure}[t]{0.2\textwidth}
        \centering
        \includegraphics[width=\textwidth]{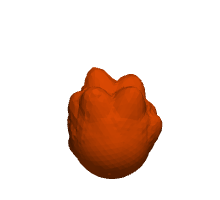}
        \caption{$4$ (default)}
        \label{fig:chindapol-n-2-2-2}
    \end{subfigure}
    \caption{Changing the \textit{sponge resolution} simulation parameter. The sponge resolution is the subdivision level of the initial icosphere sponge mesh.}
    \label{fig:sponge-res}
\end{figure*}

\begin{figure*}[t!]
    \centering
    \begin{subfigure}[t]{0.2\textwidth}
        \centering
        \includegraphics[width=\textwidth]{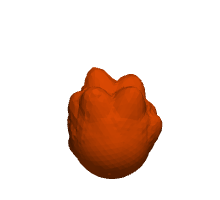}
        \caption{$0$ (default)}
        \label{fig:chindapol-n-2-2-2}
    \end{subfigure}
    \begin{subfigure}[t]{0.2\textwidth}
        \centering
        \includegraphics[width=\textwidth]{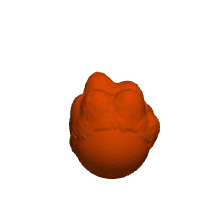}
        \caption{$0.03$}
        \label{fig:chindapol-n-1-0-2}
    \end{subfigure}
    \begin{subfigure}[t]{0.2\textwidth}
        \centering
        \includegraphics[width=\textwidth]{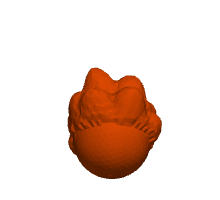}
        \caption{$0.06$}
        \label{fig:chindapol-n-1-4-2}
    \end{subfigure}
    \begin{subfigure}[t]{0.2\textwidth}
        \centering
        \includegraphics[width=\textwidth]{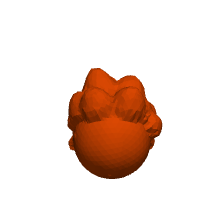}
        \caption{$0.09$}
        \label{fig:chindapol-n-1-8-2}
    \end{subfigure}
    \caption{Changing the \textit{nutrient growth threshold} simulation parameter. The sponge will only grow if the growth distance of the vertex is larger than the nutrient growth threshold value.}
    \label{fig:nut-thresh}
\end{figure*}

\begin{figure*}[t!]
    \centering
    \begin{subfigure}[t]{0.2\textwidth}
        \centering
        \includegraphics[width=\textwidth]{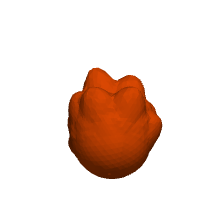}
        \caption{$0.015$}
        \label{fig:chindapol-n-2-2-2}
    \end{subfigure}
    \begin{subfigure}[t]{0.2\textwidth}
        \centering
        \includegraphics[width=\textwidth]{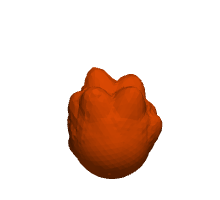}
        \caption{$0.02$ (default)}
        \label{fig:chindapol-n-1-0-2}
    \end{subfigure}
    \begin{subfigure}[t]{0.2\textwidth}
        \centering
        \includegraphics[width=\textwidth]{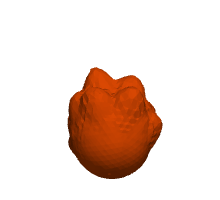}
        \caption{$0.025$}
        \label{fig:chindapol-n-1-4-2}
    \end{subfigure}
    \begin{subfigure}[t]{0.2\textwidth}
        \centering
        \includegraphics[width=\textwidth]{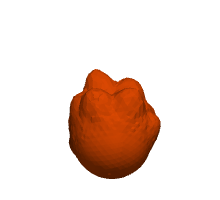}
        \caption{$0.03$}
        \label{fig:chindapol-n-1-8-2}
    \end{subfigure}
    \caption{Changing the \textit{normal offset} simulation parameter. This controls the maximum offset value when randomly offsetting the vertex normals.}
    \label{fig:norm-offset}
\end{figure*}

\begin{figure*}
    \centering
    \includegraphics[width=0.9\columnwidth]{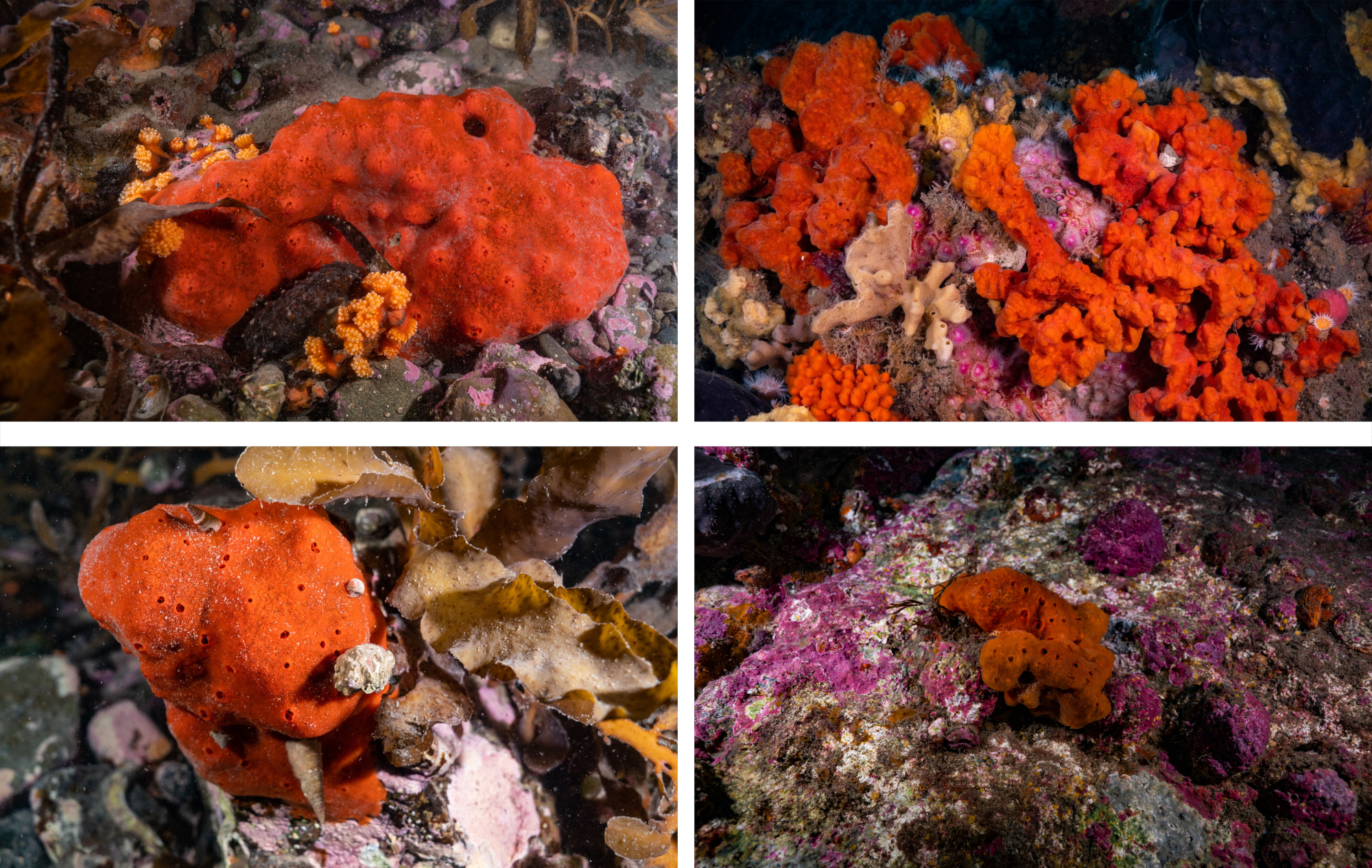}
    \caption{Photographs of the marine sponge Crella incrustans.}
    \label{fig:crella-photos}
\end{figure*}

Below we present additional results (all at $50$ growth iterations) by showing how changing simulation parameters can produce a large variation in the geometric shape of the sponge. All evaluations used the default simulation parameter values (Table \ref{table:1})  with 50 growth iterations, except for the parameters being changed. See the results in Figures~\ref{fig:c-offset}, \ref{fig:chin-n}, \ref{fig:chin-K}, \ref{fig:growth-rate}, \ref{fig:sponge-res}, \ref{fig:nut-thresh}, \ref{fig:norm-offset}. We also show additional photo comparisons in Figure~\ref{fig:comp-photos} and reference photos in Figure~\ref{fig:crella-photos}.

\begin{table*}
    \centering
    \resizebox{\textwidth}{!}{
        \begin{tabular}{|c|c|c|}
        \hline
        \textbf{Name} & \textbf{Description} & \textbf{Default Value} \\
        \hline
        \textit{diffusion limited} & If set to true, the velocity vector & True \\ 
        & is set to 0 in the advection-diffusion equation. & \\
        \hline
        \textit{concentration offset distance} & How far along the normal concentration values & $0.1$ \\
        & should be interpolated. & \\
        \hline
        \textit{box resolution} & Determines how many times to subdivide the & $2$ \\
        & FEM box mesh, with 1 being no subdivision. & \\
        & More subdivisions means more points of the & \\
        & mesh where the nutrient concentration is solved for. & \\
        \hline
        \textit{sponge resolution} & The subdivision level of the initial sponge & $4$ \\
        & icosphere mesh. More subdivisions means & ($642$ triangles) \\
        & a more accurate growth pattern. & \\
        \hline
        \textit{growth rate} & A float which scales the distance a sponge vertex & $5$ \\
        & grows in the direction of the normal. & \\
        \hline
        \textit{growth threshold} & The growth length needed for the vertex & $0$ \\
        & to actually grow. & \\
        \hline
        \textit{vertex insertion threshold} & If the distance between two vertices is & $0.6$ \\
        & larger than this number, a vertex is inserted between them. & \\
        \hline
        \textit{vertex fusion threshold} & If the distance between two vertices is  & $0.5$ \\
        & lower than this number, the two vertices are fused into one. & \\
        \hline
        \textit{triangle area deletion threshold} & If the area of a triangle is  & $0.1$ \\
        & smaller than this value, the triangle is deleted by fusing & \\
        & together the two vertices of the shortest side. & \\
        \hline
        \textit{Chindapol kinetic order} & The kinetic order of the growth rate with respect to $C_i$ & $1.4$ \\
        & in the Chindapol growth equation. & \\
        \hline
        \textit{Chindapol characteristic growth constant} & A constant which influences the characteristic growth curve & $1$ \\
        & in the Chindapol growth equation. & \\
        \hline
        \textit{skeletal growth type} & Either Radiate accretive & Radiate \\
        & (which is present in the sponge Haliclona occulata), or & accretive \\
        & Halichondrid (which is present in the sponge Crella incrustans). & \\
        \hline
        \textit{normal offset radius} & The length of the random displacement & $0.02$ \\
        & vector applied to the normal. & \\
        \hline
        \textit{max nvs iterations} & The maximum number of Navier Stokes non-linear solver iterations & $100$ \\
        & where more iterations means more & \\
        & accuracy in the fluid simulation results. & \\
        \hline
        \end{tabular}
    }
    \caption{The main simulation parameters and their default values.}
    \label{table:1}
\end{table*}

\newcommand{\vrulesep}{\unskip\ \vrule\ }
\newcommand{\hrulesep}{\unskip\ \hrule\ }

\begin{figure*}[t!]
    \centering
    \hspace*{\fill}%
    \begin{subfigure}[t]{0.45\textwidth}
        \centering
        \includegraphics[width=\textwidth]{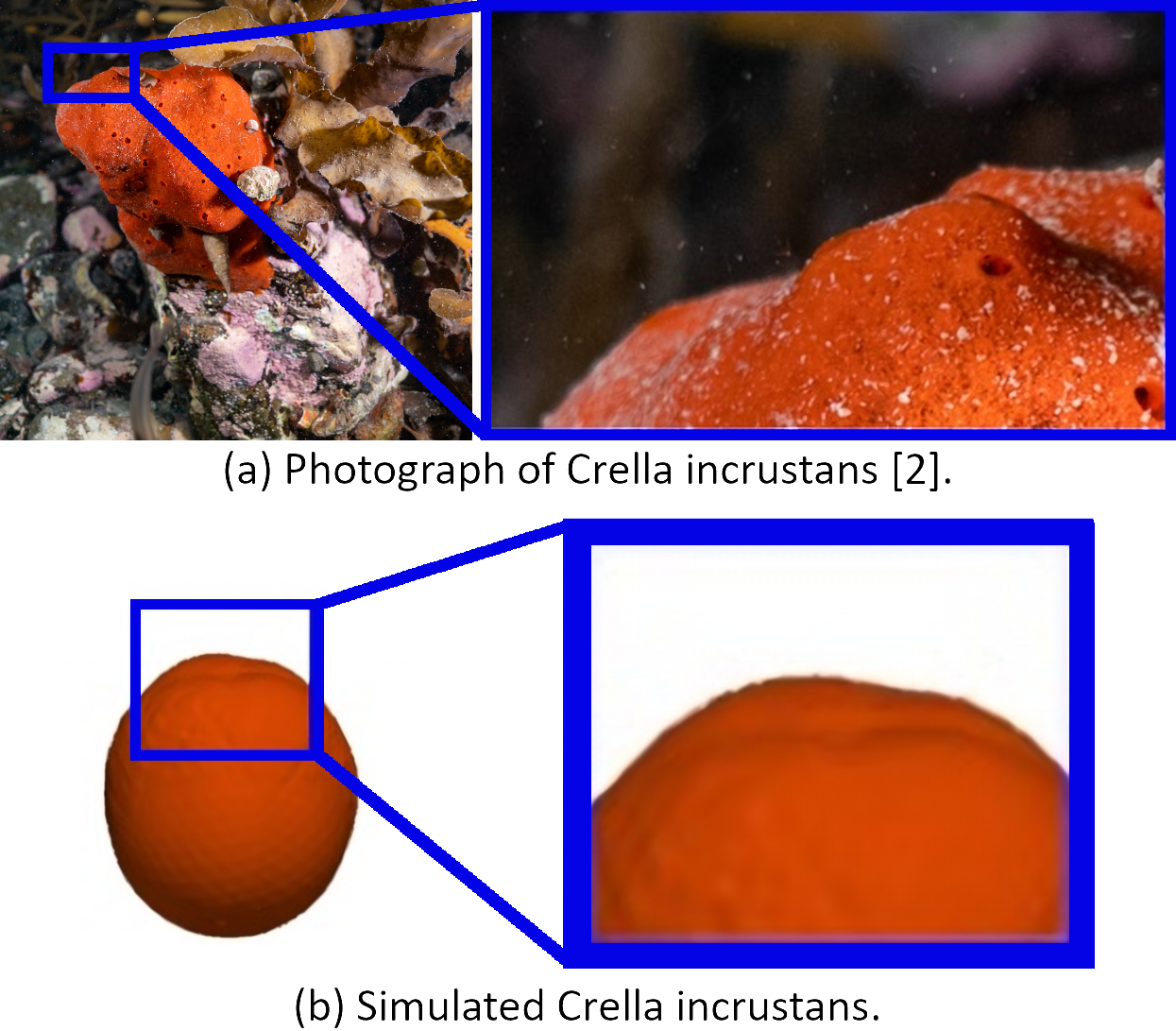}
    \end{subfigure}
    \hfill \vrule\ \hfill%
    \begin{subfigure}[t]{0.45\textwidth}
        \centering
        \includegraphics[width=\textwidth]{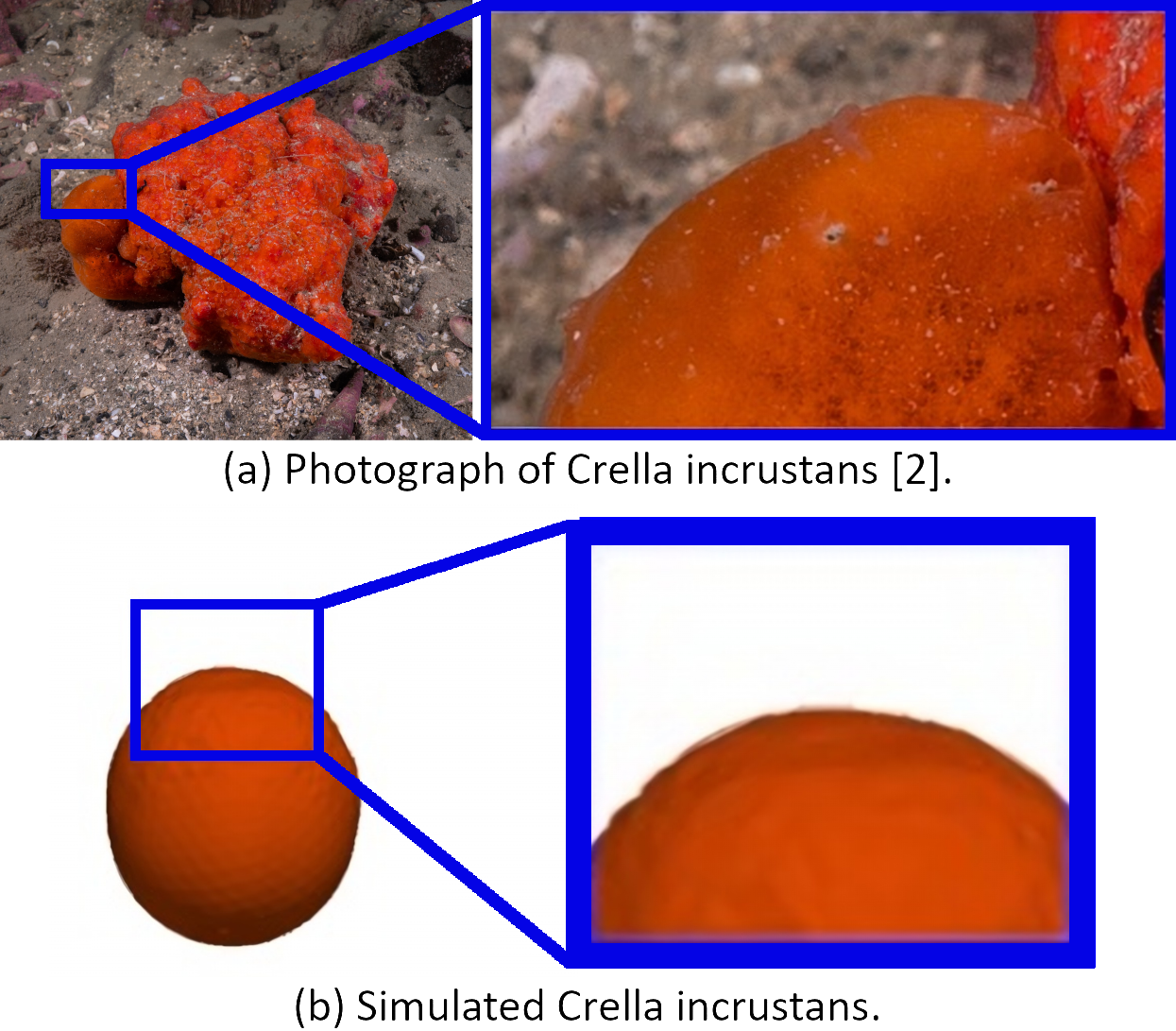}
    \end{subfigure}
    \hspace*{\fill}%

    \hrulesep
    \vspace*{0.3cm} 

    \hspace*{\fill}%
    \begin{subfigure}[t]{0.45\textwidth}
        \centering
        \includegraphics[width=\textwidth]{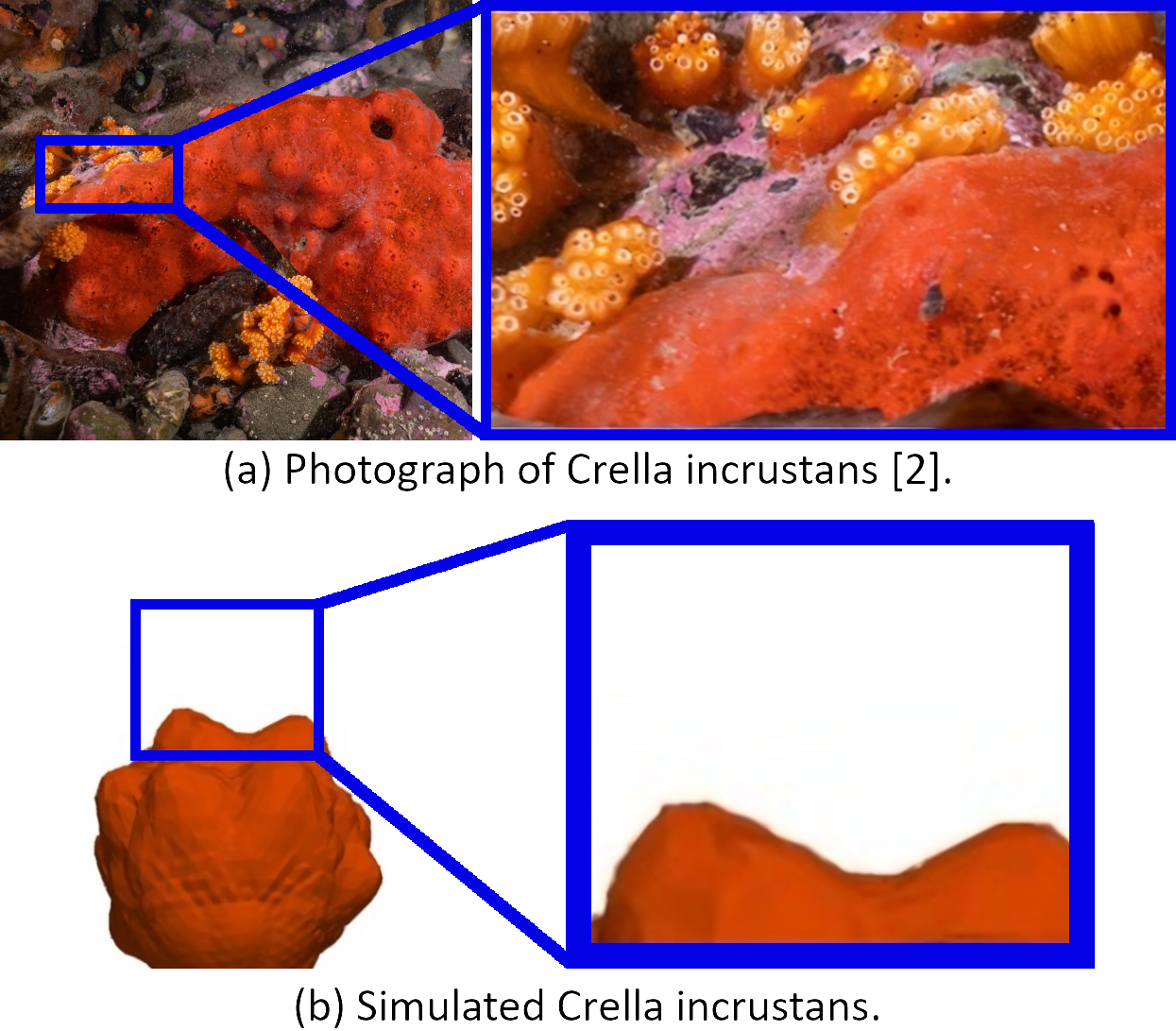}
    \end{subfigure}
    \hfill \vrule\ \hfill%
    \begin{subfigure}[t]{0.45\textwidth}
        \centering
        \includegraphics[width=\textwidth]{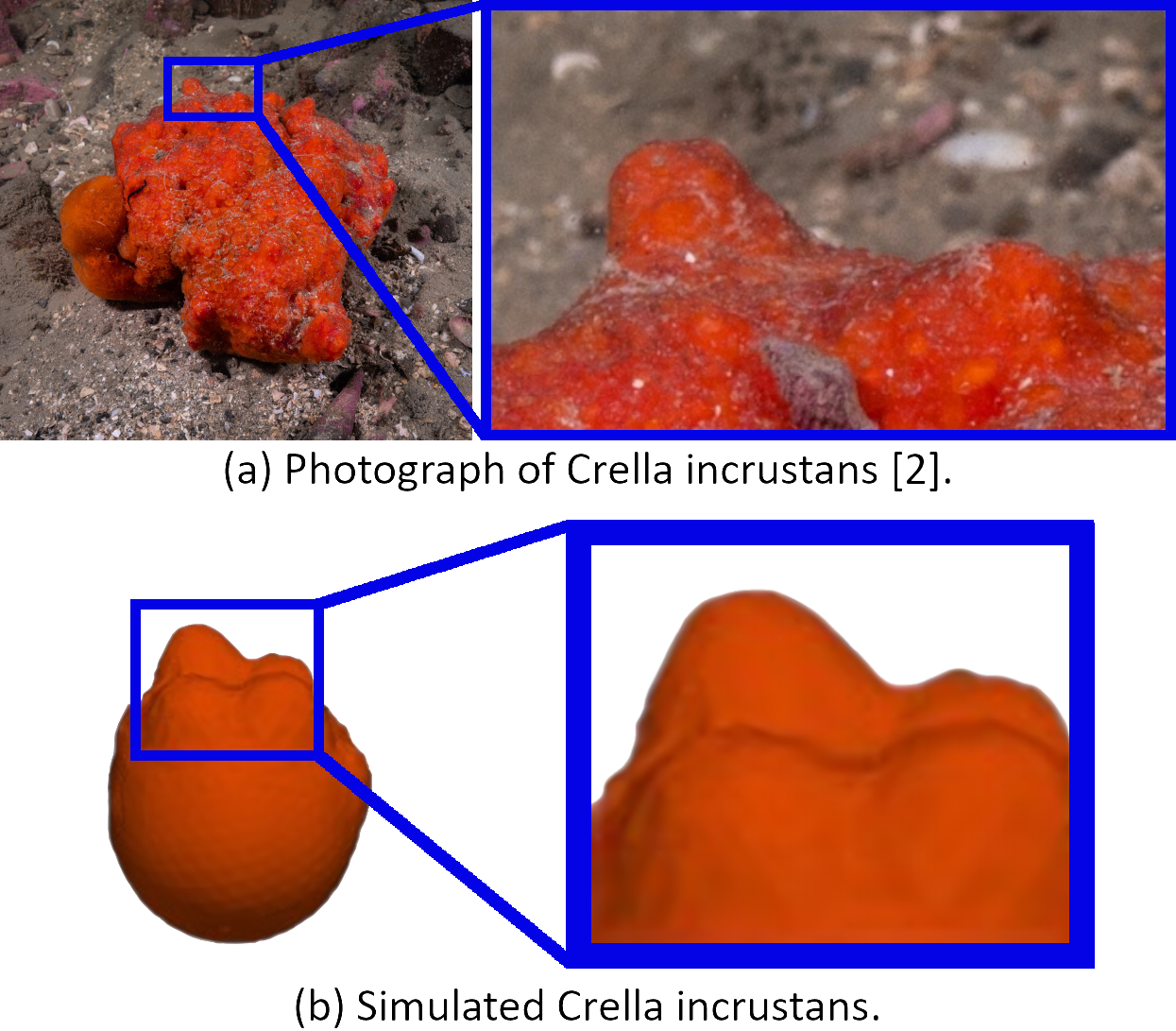}
    \end{subfigure}
    \hspace*{\fill}%

    \caption{Four examples comparing (a) a real-life photo with (b) the simulated Crella incrustans sponge.}
    \label{fig:comp-photos}
\end{figure*}


